\documentclass{article}
\usepackage{PRIMEarxiv}

\usepackage[utf8]{inputenc} 
\usepackage[T1]{fontenc}    
\usepackage{hyperref}       
\usepackage{url}            
\usepackage{booktabs}       
\usepackage{amsfonts}       
\usepackage{nicefrac}       
\usepackage{microtype}      
\usepackage{amsmath} 
\usepackage{cleveref}       
\usepackage{lipsum}         
\usepackage{graphicx}
\usepackage{doi}

\usepackage{longtable}
\usepackage{float}
\usepackage{multirow}
\usepackage{makecell}
\usepackage{amsmath}
\usepackage{lscape}
\newcommand*\samethanks[1][\value{footnote}]{\footnotemark[#1]}
\title{A renewable power system for an off-grid sustainable telescope fueled by solar power, batteries and green hydrogen
}

\author{
  Isabelle Viole\thanks{These authors contributed equally to this work.} , Guillermo Valenzuela-Venegas\samethanks[1], Marianne Zeyringer and Sabrina Sartori \\
  Department of Technology Systems (ITS)\\
  University of Oslo\\
  Kjeller, Norway \\
  \texttt{isabelle.viole@its.uio.no(IV); guillermo.valenzuela@its.uio.no (GVV)} }

\begin{document}
\maketitle

\begin{abstract}
A large portion of astronomy’s carbon footprint stems from fossil fuels supplying the power demand of astronomical observatories. Here, we explore various isolated low-carbon power system setups for the newly planned Atacama Large Aperture Submillimeter Telescope, and compare them to a business-as-usual diesel power generated system. Technologies included in the designed systems are photovoltaics, concentrated solar power, diesel generators, batteries, and hydrogen storage. We adapt the electricity system optimization model highRES to this case study and feed it with the telescope’s projected energy demand, cost assumptions for the year 2030 and site-specific capacity factors. Our results show that the lowest-cost system with LCOEs of \$116/MWh majorly uses photovoltaics paired with batteries and fuel cells running on imported and on-site produced green hydrogen. Some diesel generators run for backup. This solution would reduce the telescope’s power-side carbon footprint by 95\% compared to the business-as-usual case.
\end{abstract}


\section{Introduction}
\label{sec1}

To contribute to the global effort limiting the global rise in temperature below 1.5$^{\circ}$, astronomers need to reduce their carbon footprint. Today’s carbon footprint per astronomer is estimated at \textasciitilde37t CO$_{2}$ equivalents (CO$_{2}$e) annually \cite{stevens_imperative_2020, knodlseder_estimate_2022}. The sector as a whole emits \textasciitilde1,200kt CO$_{2}$e per annum, that is 0.0024\% of the global greenhouse gas (GHG) emissions (50.2Gt in 2020). This exceeds the population share of this group by more than a factor of six, since just 30,000 out of the current population of 8bn are astronomers \cite{knodlseder_estimate_2022}. Next to CO$_{2}$e emissions related to flights and powering supercomputers, 13\% stems from powering the operation of observatories \cite{stevens_imperative_2020}. These have high power needs for the cryogenic cooling of instruments and the motors tracking their dishes. As most of this electricity stems from fossil fuel sources today, ground-based telescopes have a median carbon intensity of 24t CO$_{2}$e per paper published using their generated data \cite{knodlseder_estimate_2022}. For comparison, the carbon footprint per capita in the European Union was 6.8t CO$_{2}$e in the year 2019 \cite{european_commission_statistical_office_of_the_european_union_energy_2020}.

Growing awareness of climate change mitigation and volatile prices of fossil fuels led some observatories to subsequently add Renewable Energy Sources (RES) to their power generation. In Hawaii, the Gemini Observatory supplies 10\% of its demand with solar photovoltaic (PV) arrays since 2015. In the Chilean Atacama Desert, the La Silla observatory is powered by more than 50\% solar energy since 2016 and the Paranal observatory commissioned PV arrays in 2022 \cite{acohido_gemini_2017, de_zeeuw_reaching_2016, francisco_rodriguez_esos_2022}.

Due to the decades-long lifetime of astronomical research infrastructures, decisions taken on power supply in new projects today lock in CO$_{2}$e emissions for decades. To strive for lower carbon footprints, AtLAST - The Atacama Large Aperture Submillimeter Telescope - is the first observatory already including plans for a power system powered by RES in its design stage \cite{klaassen_atacama_2020}. AtLAST is a design study project for a 50m single dish submillimeter/millimeter telescope on the Chajnator plateau in the Atacama, at an elevation of \textasciitilde5,000 meters \cite{klaassen_atacama_2020, ramasawmy_atacama_2022}. It is planned to start its operation in the early 2030s. The very high and dry site is necessary to minimize water vapor in the atmosphere, which absorbs submillimeter photons. The eleven neighboring telescopes, like the Atacama Pathfinder Experiment (APEX) and the Atacama Large Millimeter/ submillimeter Array (ALMA), today rely on fossil fuels to meet their power demands.

Several recent studies carried out simulations and techno-economic assessments of islanded stationary systems for household demands lately, considering present-day component costs. Endo et al. \cite{endo_simulation_2019} simulated the operation of a stationary system to power a residential building, consisting of RES, an electrolyzer, a hydrogen (H$_2$) storage system in metal hydrides (MH), a proton-exchange membrane fuel cell (PEMFC) and batteries. They showed that PV and PEMFC generation can meet most of the demand during fine and cloudy weather. Ghenai et al. techno-economically optimized an off-grid solar PV and H$_{2}$ energy system to meet the 500kWh daily demand of a desert community, resulting in levelized costs of electricity (LCOE) of \$145/MWh \cite{ghenai_technico-economic_2020}. Gebrehiwot et al. similarly analyzed a hybrid power system with PV, wind, and diesel generation plus batteries to cover a remote village's demand, LCOEs resulting in \$207/MWh \cite{gebrehiwot_optimization_2019}. Dawood et al. compared off-grid systems with PV, batteries and/or H$_{2}$ storage for a remote settlement demanding 2MWh daily, where a hybrid energy storage system with both batteries and H$_{2}$ induced lowest costs of electricity with \$342/MWh. This more than halved the PV curtailment compared to coupling PV only with batteries \cite{dawood_stand-alone_2020}.

We identify a number of research gaps in literature: First, to our knowledge, there is no similar study analyzing the design of a renewable energy system for a remote research facility in general and telescope in particular. Due to very different needs as well as environmental conditions study results from residential settings can't be applied to telescopes. Further, studies on islanded, renewable-based energy systems have not been applied to high altitudes yet, only use a limited number of technologies and do not consider technology learning. 

In this work we close these gaps by: 
1. performing the first optimization of a power system for a remote telescope with a unique electricity demand, characterized by seasonal shifts and minimal day-night fluctuations. It serves as a pioneering example for the transition to RES in off-grid research facilities, crucial for reducing CO$_{2}$ emissions to meet climate targets. 
2. considering elevation-specific derating factors serving as a first study in the design of high-altitude power systems 
3. providing a more comprehensive comparison of system setups based on renewable energy for remote and off-grid power systems: We evaluate the trade-offs between 100\% renewable systems, and systems with fossil fuel backup, next to trade-offs between batteries and hybrid energy storage systems, offering insights into the most cost-effective and sustainable solutions. We compare systems using imported and on-site produced green H$_{2}$ to those only employing batteries. Concentrated Solar Power (CSP), a promising technology in high irradiation places, is compared to the more widely applied PV technology. 
4. applying technology learning rates in a sensitivity analysis: In line with the projected start of AtLAST's operations in the early 2030s, we size a power system for 2030 with a wide range of cost decline assumptions in RES and varying fuel costs. In contrast to modeling the system with today's costs, the use of projected future costs paired with a cost-sensitivity analysis helps to better showcase relevant energy systems for AtLAST in the next decade in which it will need to be supplied. It thereby also represents a valuable case study for similar, remote infrastructure projects that are currently being planned.

Specifically, this work presents solutions to power AtLAST’s consumption of annually \textasciitilde7.7GWh with an isolated power system, including business-as-usual powering with fossil fuel generation, and combinations of RES, energy storage and fossil fuel generation backup. We answer the following research questions:
\begin{enumerate}
\item Which system results in the lowest levelized costs of electricity (LCOE) to power the planned telescope?
\item How much more does it cost to use a power system without direct CO$_2$e-emissions?
\item	Which system setups are robust against uncertainty in future costs?
\end{enumerate}

\section{Materials and Methods}
\label{sec2}
We apply a cost-optimization model to find feasible power systems for the telescope, divided into eight scenarios:
\begin{enumerate}
\item Using diesel generators at 5,000m (Business-as-usual, BAU); 
\item Combining PV and diesel generation (PV \& Diesel, PVD) 
\begin{itemize}
\item at 5,000m (PVD↑); 
\item at 2,500m (PVD↓); 
\end{itemize}
\item Combining PV generation with a hybrid energy storage system
\begin{enumerate}
\item	with backup diesel generation (PV, Diesel \& Energy Storage, PVDES)
\begin{itemize}
\item at 5,000m (PVDED↑); 
\item at 2,500m (PVDES↓); 
\end{itemize}
\item as a renewables-only system without diesel generation (Renewable Energy Sources, RES)
\begin{itemize}
\item at 5,000m (RES↑); 
\item at 2,500m (RES↓); 
\end{itemize}
\end{enumerate}
\item Using CSP at an altitude of 2,500m (CSP).
\end{enumerate}

\begin{figure}
\centering
\includegraphics[width=10.5cm]{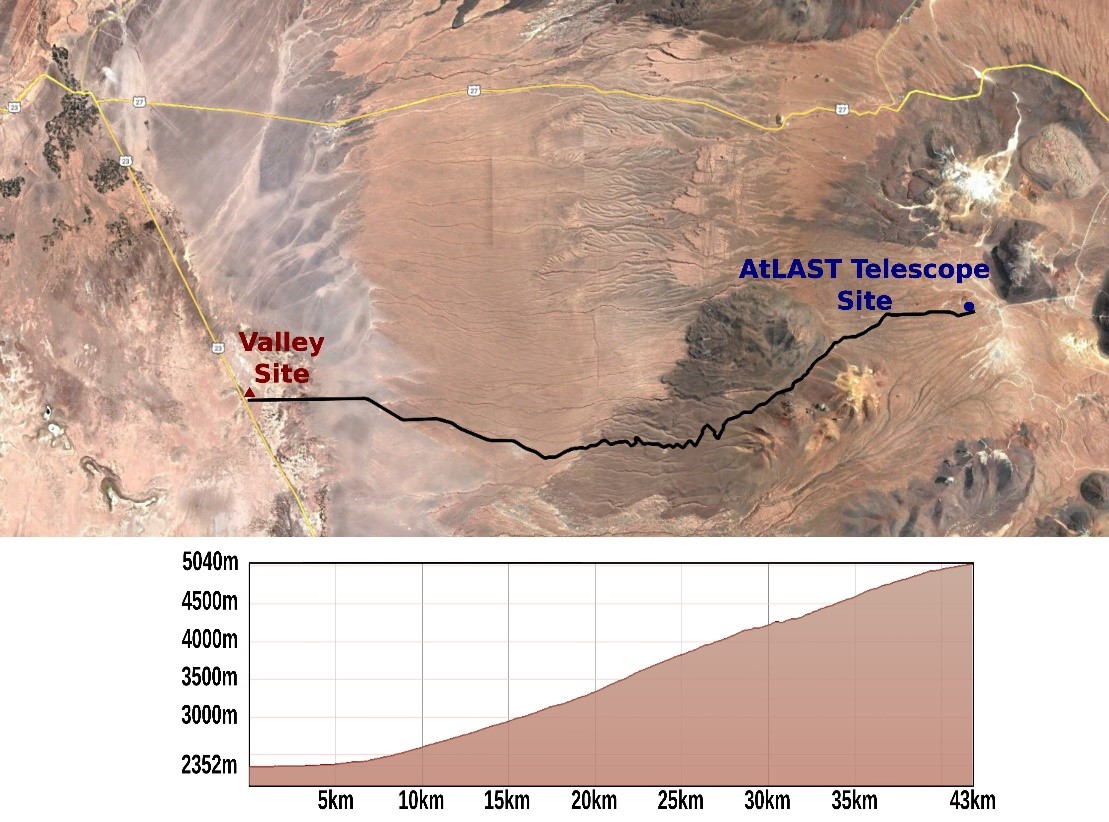}
\caption{Map of the studied area, highlighting a potential sites of AtLAST, the power generation location in the valley and the necessary power line pathway between the two.\label{fig.sites}}
\end{figure}

2030 is set as the potential year to start operation. Location-wise, we consider systems built near the telescope on the Chajnator plateau at \textasciitilde5,000m and system at a Valley Site, 2,500m above sea level, see Figure \ref{fig.sites}. For the telescope's location, currently two options less than 5km apart from each other are investigated. These are undistinguishable irradiation-wise, both present temperature variations from -20 to 10\textdegree C and low pressures of \textasciitilde550Pa, for which we apply derating factors, and hence are considered alike in this work. The Valley Site, 43km away from the telescope, implies less derating of components and easier access for operations, while adding costs for subterranean cabling. Only off-grid systems are considered.

\begin{figure}
\centering
\includegraphics[width=10.5cm]{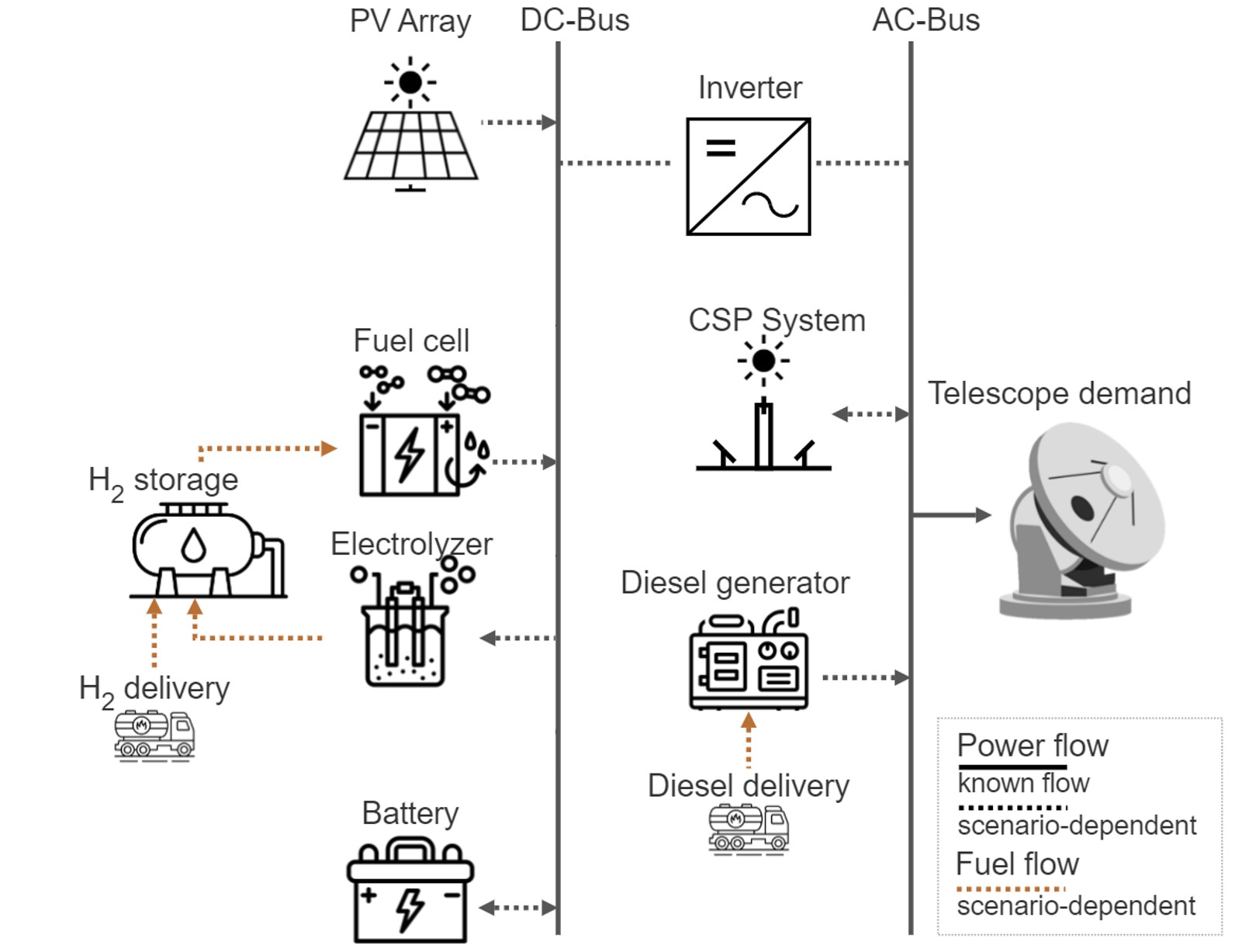}
\caption{Potential components of power system for the telescope, inverter not included in optimization.\label{fig.setup}}
\end{figure}

To optimize the different scenarios, we apply highRES-AtLAST, an adaptation of the highRES model proposed by Price and Zeyringer for the energy system of Great Britain and Europe \cite{price_highres-europe_2022, zeyringer_designing_2018}. Constructed to design a cost-minimal power system using a linear programming model, it minimizes overall costs while meeting hourly demands and technical constraints. The main outputs are system costs, built capacities and hourly dispatch. We add a H$_{2}$ storage system with electrolyzers, storage and fuel cells and CSP as technologies to the model. The H$_{2}$ and CSP system (thermal energy) were implemented in terms of power (MW and MWh) to fit with the highRES limitation. The limitations of highRES-AtLAST are the same as the original model. Figure \ref{fig.setup} shows an overview of all technologies in the system. 2030 serves as the demand and cost assumption year, while costs are set to inflation-adjusted real2022-US\$ \cite{international_monetary_fund_world_nodate}. Inputs for the model include solar generation capacity factors, Chilean component and fuel costs in 2030 and a demand time series for AtLAST. Different learning rates for component costs and variations in fuel costs are integrated via sensitivity analyses.

\subsection{Estimating the telescope’s demand}
\label{sec2-3}
\begin{figure}
\centering
\includegraphics[width=16.5cm]{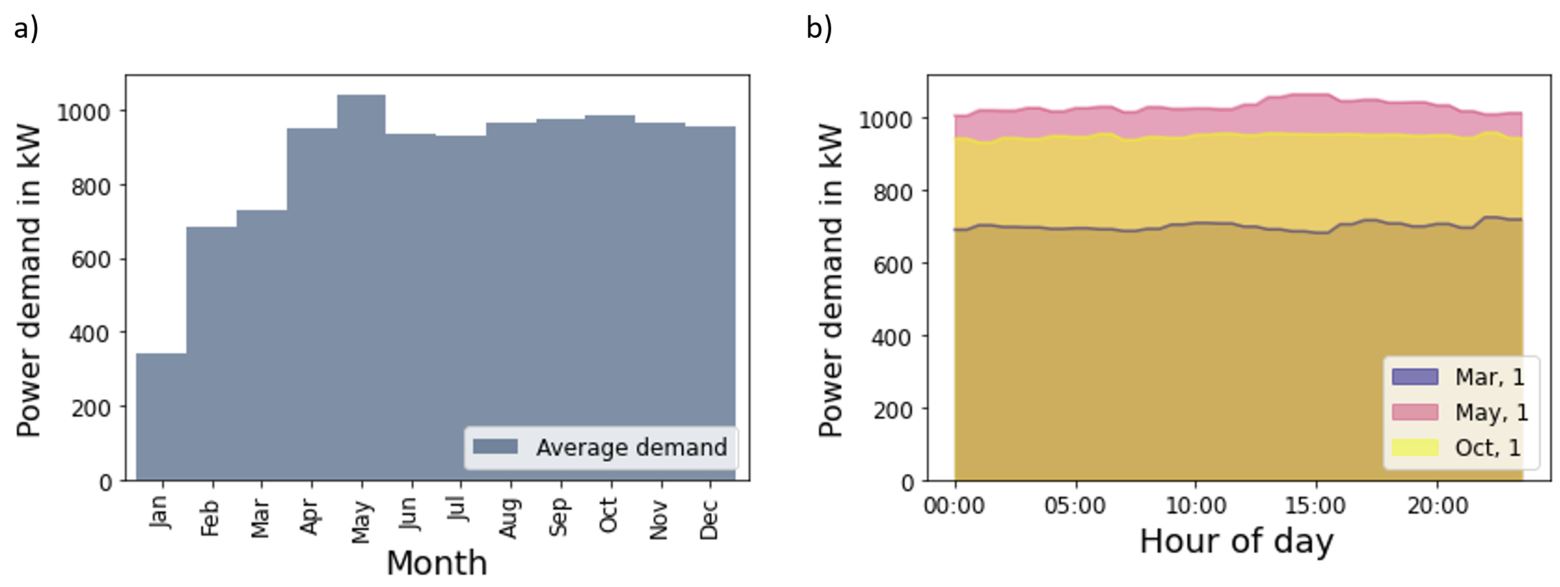}
\caption{Demand patterns of AtLAST in 2030, a) average monthly demand; b) exemplary daily demand curves.\label{fig.atldem}}
\end{figure}

The demand of AtLAST comprises dish motors, a cryogenic cooling system for the instruments, electronics, instrument support, and a heating, ventilation, air conditioning system, see Table \ref{t.dem}. Peak demands are supplied by supercapacitors outside the system presented. For the optimization, the historical demand of a smaller radio telescope is upscaled to match AtLAST's estimated demand, resulting in a one-year time series for 2030 at hourly granularity. Figure \ref{fig.atldem}a) shows the estimated average monthly power demand, see Figure \ref{fig.atldem}b) for exemplary daily demand curves.

\begin{table}
\caption{Demand estimation of AtLAST.\label{t.dem}}
\begin{tabular}{lrr}
\toprule
Demand constituent	& Average demand (kW)	& Peak demand (kW)\\
\midrule
Cryogenic cooling&450&600\\
Telescope motors&350&1300\\
Heating, ventilation and air conditioning system&40&50\\
Electronics; instrument support; other&40&50\\
\bottomrule
\end{tabular}
\end{table}

\subsection{Choice of system components}
\label{sec2-1}
\begin{figure}
\centering
\includegraphics[width=16.5cm]{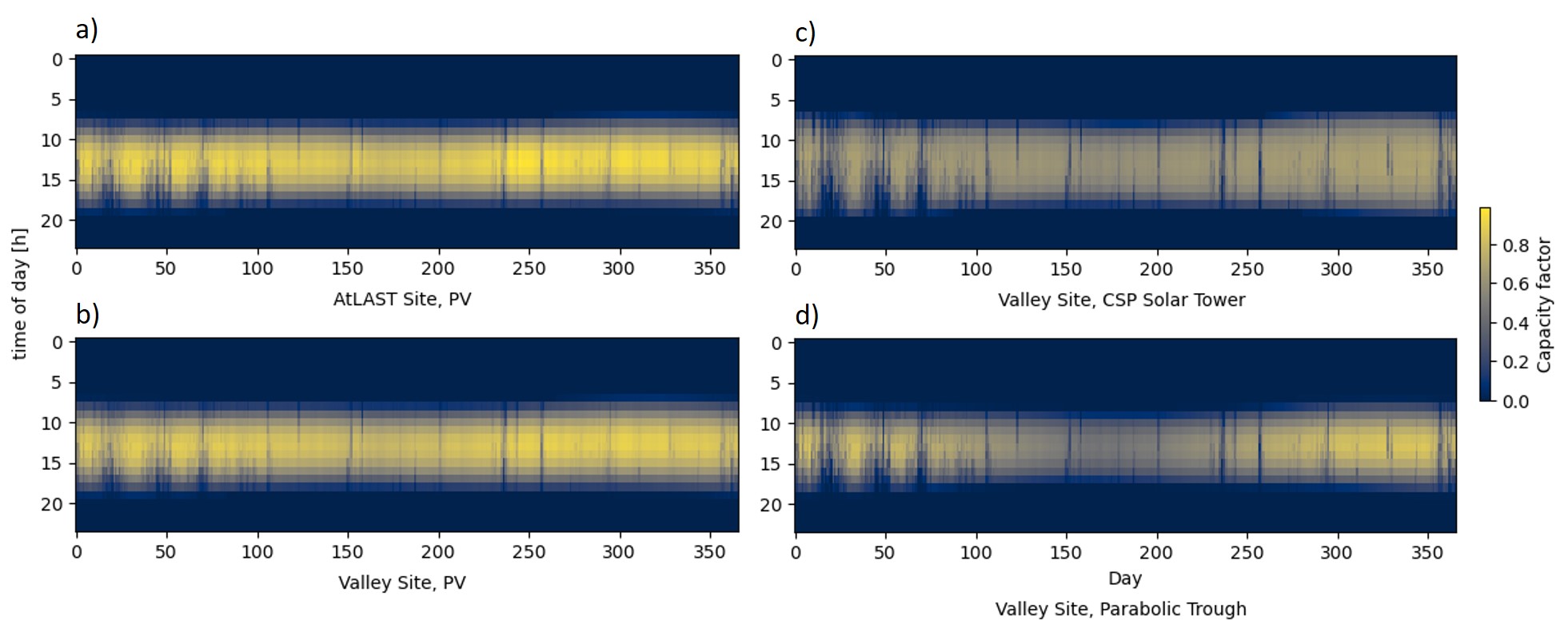}
\caption{Solar generation capacity factors: (a) PV at AtLAST Site, (b) at Valley Site, (c) CSP ST at Valley Site, (d) CSP PT at Valley Site.\label{fig.capa}}
\end{figure}

The global solar irradiation maximum lies within the Atacama Desert, an annual irradiation of \textasciitilde2,640kWh/m$^{2}$ can be assumed at the prospective sites \cite{rondanelli_atacama_2015, hersbach_era5_2020}. As inputs for the optimization, we convert hourly irradiation values from ERA5, previously used to represent the weather conditions on a regional level in modelling the Chilean electricity generation \cite{ramirez_camargo_assessment_2019}, into capacity factors \cite{hersbach_era5_2020, hofmann_atlite_2021}, see Figure \ref{fig.capa}. While PV systems can be located at either 5,000m or 2,500m, CSP is excluded at the high site, to not interfere with astronomical observations. Wind power is not considered, as the areal average wind speed of 3.1m/s is too low for power production \cite{munoz_wind_2018}.

To represent CSP in the model we consider three interlinked subsystems: the solar field (capturing solar irradiation), the thermal energy storage (storing surplus thermal energy), and the power block (generating power) \cite{chattopadhyay_capacity_2018}. We consider the two most established CSP technologies, the Parabolic Trough (PT) and the Solar Tower (ST) \cite{raboaca_concentrating_2019}.
Advantages lie in their flexibility, as thermal energy storage allows for power generation on demand. However, CSP is often not competitive with the steeply declining PV costs, so some systems have been retrofitted to PV \cite{schoniger_making_2021}. 

Today’s telescopes in the area are for the majority supplied by diesel generators, which can ramp up and down flexibly and are dependent on regular refueling trucks. We employ similar diesel generators in our model.

\subsubsection{Energy storage systems}
\label{sec2-1-2}
We consider both batteries and H$_{2}$ as energy storage technologies. In batteries, mature options include lead acid and lithium-ion (Li-ion) batteries. According to manufacturers, both technologies can withstand temperatures down to -20\textdegree C in discharge/storage, and down to 0\textdegree C when charging. Lead acid batteries with their historically lower investment costs currently hold the leading role in stationary systems. In recent years, however, some found lower LCOEs with Li-ion batteries, based on their higher cycle life \cite{kebede_techno-economic_2021}. As additionally, Li-ion batteries perform better in life cycle assessments when compared to lead acid, we include only this technology in our setup \cite{yudhistira_comparative_2022}.

When utilizing H$_{2}$ as an energy carrier in remote systems, two ways of operation are thinkable. For one, H$_{2}$ production off-site and shipment to the location is doable, requiring on-site storage and fuel cells to generate power. We only consider so-called green H$_{2}$, stemming from RES-powered electrolysis. Without imported H$_{2}$, the whole H$_{2}$ value chain is needed on-site, that is electrolyzers, H$_{2}$ storage and fuel cells. H$_{2}$ storage in stationary systems today is mostly conducted with compressed gas (CG) stored in steel cylinders at 50-700 bar \cite{egeland-eriksen_hydrogen-based_2021}. MH applications for storing H$_{2}$ are investigated as a safer alternative to CG, e.g. by Endo et al. \cite{endo_simulation_2019}, but are not economically available yet.

\subsubsection{Derating factors}
\label{sec2-1-4}
We do not know of any solar power system with a substantial energy storage system as proposed in this work in existence at 5,000m. In this techno-economical assessment, we consider technologies that could endure the harsh conditions of the location, and include derating factors to consider the low-pressure conditions at either site, see Table \ref{t.derating}.
These factors impact the components' power output. For instance, a diesel generator with a capacity of 2MW at the ATLAST Site has a maximum effective capacity of 1MW, that is its capacity multiplied with a derating factor of 0.5. Diesel generators are impacted by the less dense air, as less oxygen molecules are available for the burning process. Batteries are derated at altitudes, as the extremely low temperatures and low pressures represent an issue for their liquid electrolyte. Both electrolyzers and fuel cells are impacted by the lower amount of oxygen molecules available. The factors of diesel generators and batteries stem from empirical data and internal communications with experts from other telescopes in the areas. In experiments of Pratt et al., a PEMFC at an altitude of 2,774 meters had 6-10\% less stack voltage than its counterpart at an altitude of zero \cite{pratt_performance_2007}. We hence assume similar derating factors for H$_{2}$ and batteries.

\begin{table}
\caption{Derating factors at 2,500 and 5,000 altitude meters.\label{t.derating}}
\begin{tabular}{lrr}
\toprule
System component	& Derating factor at 2,500m	& Derating factor at 5,000m\\
\midrule
Diesel generator&0.75&0.5\\
Electrolyzer/Fuel Cell/Li-ion battery &0.95&0.8\\
\bottomrule
\end{tabular}
\end{table}

\subsection{Component and fuel costs, sensitivity analysis}
\label{sec2-2}

PV, batteries, and H$_{2}$ systems had a steep learning curve over the last decades, a trend forecasted to continue until beyond 2030, the planned building year of this system. In this work, we use Chilean forecasted costs, wherever possible, and integrate different learning rates in sensitivity analyses to account for variations in future component costs. The base case capital expenditures (CAPEX) and lifetime assumptions are shown in Table \ref{t.capex}, with costs adjusted to real2022-US\$ values. See Table \ref{ta.capexopex} in Appendix for operational expenditures (OPEX). This optimization includes generation, storage units and long-distance cabling as the major cost components. Other electrical system components like inverters are not integrated, though they need to be considered in any final system design. Sensitivity analyses with costs as indicated in Table \ref{t.sens} and Table \ref{t.fuelcosts} are performed to indicate the robustness of system setups against different learning curves and other fuel costs than the ones assumed in the base scenario.

\begin{table}
\caption{Cost estimations in 2030, real2022-US\$ values.\label{t.capex}}
\begin{tabular}{lrrr }
\toprule
Cost component	& CAPEX	& Unit & Lifetime\\
\midrule
Monofacial monocrystalline PV & 523 \cite{nama_facility_indice_2020, vartiainen_impact_2020} &
\$/kW$_{\mathrm{p}}$ &\multirow{5}{*}{25 years} \\
CSP solar field – PT/ST & 209/148 \cite{turchi_csp_2019}&
\$/m$^{2}$ \\
CSP thermal energy storage – PT/ST & 37/19 \cite{schoniger_making_2021, turchi_csp_2019, dieckmann_lcoe_2017} &
\$/kWh$_{\mathrm{th}}$\\
CSP power block – PT/ST & 869/1,156 \cite{turchi_csp_2019} & \$/kW$_{\mathrm{e}}$\\
Diesel generator & 495 \cite{comision_nacional_de_energia_informe_2020}
&\$/kW\\
Li-Ion batteries & 262 \cite{cole_cost_2021, mongird_2020_2020}
&\$/kWh & 13 years \cite{cole_cost_2021} \\
Electrolyzer - Alkaline / PEM & 434/483 \cite{mongird_2020_2020}
&\$/kW$_{\mathrm{e}}$ & \multirow{3}{*}{18 years \cite{cole_cost_2021}} \\
CG H$_{2}$ storage & 598 \cite{sens_conditioned_2022}
&\$/kg H$_{2}$ stored& \\
PEMFC & 1,581 \cite{cigolotti_comprehensive_2021}
&\$/kW$_{\mathrm{e}}$ & \\
Subterranean power lines, 24kV & 23,599 \cite{cype_ingenieros_sa_precio_nodate}
&\$/km & 40 years\\
\bottomrule
\end{tabular}
\end{table}

\begin{table}
\caption{Sensitivity case cost estimations in 2030, real2022-US\$ values.\label{t.sens}}
\begin{tabular}{lrrr}
\toprule
Cost component	& Low case CAPEX	& High case CAPEX & Unit\\
\midrule
Monofacial monocrystalline PV&345 \cite{nama_facility_indice_2020, vartiainen_impact_2020} &
928 \cite{international_renewable_energy_agency_future_2019}
&\$/kW$_{\mathrm{p}}$\\
Li-ion batteries&164 \cite{cole_cost_2021} &
304 \cite{mongird_2020_2020}
&\$/kWh\\
Electrolyzer - Alkaline / PEM&247/327 \cite{hydrogen_council_hydrogen_2021} &
774 \cite{international_renewable_energy_agency_future_2019}
&\$/kW$_{\mathrm{e}}$\\
CG H$_{2}$ storage & 286 \cite{sens_conditioned_2022} &
780 \cite{sens_conditioned_2022}
&\$/kg H$_{2}$ stored\\
PEMFC&1,002 \cite{clean_hydrogen_partnership_hydrogen_2019} &
1,876 \cite{cigolotti_comprehensive_2021}
&\$/kW$_{\mathrm{e}}$\\
\bottomrule
\end{tabular}
\end{table}

\begin{table}
\caption{Fuel costs in 2030, base case and high-cost sensitivity, real2022-US\$ values.\label{t.fuelcosts}}
\begin{tabular}{lrrr}
\toprule
Fuel	& Base case costs	& High case costs & Unit\\
\midrule
Diesel & 600.76 \cite{rystad_energy_research_2022} & 1,220.52 \cite{international_renewable_energy_agency_global_2022} & \$/m$^3$\\
Green H$_{2}$ & 1.13 \cite{endo_operation_2017} & 1.40 \cite{cigolotti_comprehensive_2021} & \$/kg\\
Transporting H$_{2}$ & 0.54 \cite{cabello_aprobacion_2022}& & \$/kg/100km\\
\bottomrule
\end{tabular}
\end{table}

\subsubsection{Power generation costs}
\label{sec2-2-1}
\begin{figure}
\centering
\includegraphics[width=16.5cm]{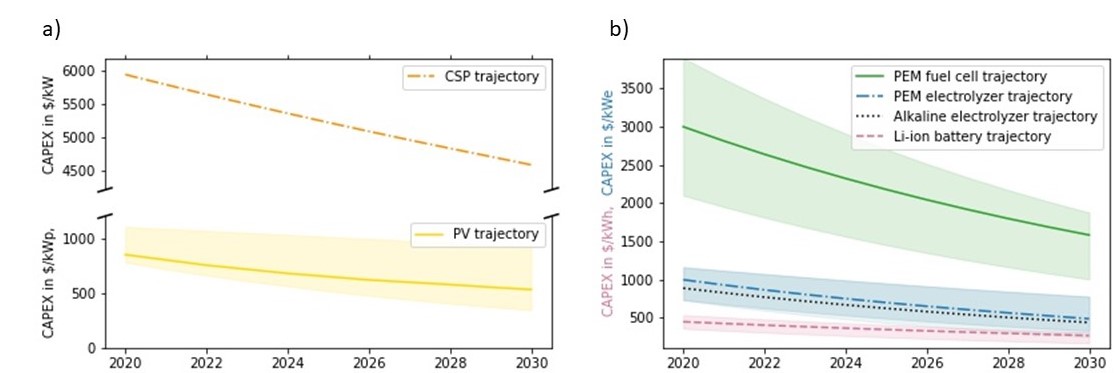}
\caption{Cost trajectory from 2020 to 2030 a) for solar power generation, PV margins for cost sensitivities; b) for energy storage components, margins for cost sensitivities.\label{fig.learnpv}}
\end{figure}

In 2020, the median costs of Chilean PV systems >1MW$_{\mathrm{p}}$ was at $\$_{2020}$769/kW$_{\mathrm{p}}$ (real2020-\$, inflation-adjusted US dollar value to the year 2020) \cite{nama_facility_indice_2020}. To forecast future costs we apply Vartiainen et al.’s base PV cost decrease rate of 39.3\% from 2020 to 2030 on the Chile-specific costs, resulting in \$515.90/kW$_{\mathrm{p}}$ in 2030 \cite{nama_facility_indice_2020, vartiainen_impact_2020}. This is comparable with the International Renewable Energy Agency (IRENA)’s (2019) assumptions of $\$_{2019}$340-834/kW$_{\mathrm{p}}$ for utility-scale PV plants in 2030 \cite{international_renewable_energy_agency_future_2019}. For sensitivity cases, we include IRENA’s slow learning rate as our high, and the fast learning rate of Vartiainen et al. with a cost decrease of 59.4\% as our low case sensitivity. See Table \ref{t.sens} for sensitivity case assumptions. Figure \ref{fig.learnpv} a) shows the cost trajectories for PV and CSP. 

With CSP, several studies report learning rates for this technology, though few present results based on future scenarios. The Joint Research Centre forecasts a learning rate of 33\% from 2015 to 2030 \cite{european_commission_joint_research_centre_cost_2018}. The Chilean Energy Ministry projects CSP cost decline of 23\% in its carbon neutrality scenario \cite{ministerio_de_energia_gobierno_de_chile_planificacion_2021}. Slower and faster learning rates range from 5-49\%. We adopt this Chilean carbon neutrality learning rate in our optimization, modularizing it on individual component costs from the National Renewable Energy Laboratory (NREL) \cite{turchi_csp_2019}. CSP system costs in this work exceed those of PV and diesel-systems by far, so no sensitivity analysis is undertaken. 

\subsubsection{Energy storage costs}
\label{sec2-2-2}
By 2030, Li-ion costs are assumed to decline from 2020's $\$_{2020}$350 to $\$_{2020}$270/kWh or $\$_{2020}$148-250/kWh by the the US-American Department of Energy (DOE) and NREL, see energy storage cost declines in Figure \ref{fig.learnpv} b) \cite{mongird_2020_2020, cole_cost_2021}. For our base case, we average between the medium NREL projections DOE’s \$304/kWh, setting the value at \$262/kWh. The lower end of NREL’s projection serves as the low and the DOE’s normal assumption as the high sensitivity.

H$_{2}$ system cost decline projections are changing rapidly. In 2019, the International Energy Agency (IEA) assumed a technology-neutral decline in electrolysis costs from $\$_{2019}$900/kW$_{\mathrm{e}}$ down to $\$_{2019}$700/kW$_{\mathrm{e}}$ in 2030 \cite{international_energy_agency_future_2019}. The DOE (2020) assumed PEM electrolyzers costs to go down to \$393-481/kW$_{\mathrm{e}}$ by 2030 \cite{mongird_2020_2020}. In 2021, the Hydrogen Council and McKinsey \& Company forecasted a decline from $\$_{2021}$550-1,050/kW$_{\mathrm{e}}$ towards $\$_{2021}$230-380/kW$_{\mathrm{e}}$ in 2030 \cite{hydrogen_council_hydrogen_2021}. Industrial electrolyzer suppliers contacted by the authors expressed that CAPEX they see with vendors beat the recent literature predictions. We use the DOE’s mean value as our base case at \$486/kW$_{\mathrm{e}}$, the Hydrogen Council’s mean value of \$327/kW$_{\mathrm{e}}$ as the low sensitivity, and the IEA’s \$774/kW$_{\mathrm{e}}$ as the high sensitivity case.

The more mature alkaline electrolyzer technology has lower investment costs than their PEM counterparts today, though Saba et al. foresee the costs of PEM systems approach those of alkaline electrolyzers soon \cite{saba_investment_2018}. We set base alkaline electrolyzer costs at the lower end of the DOE’s PEM electrolyzer cost forecasts at \$434/kW$_{\mathrm{e}}$ \cite{mongird_2020_2020}, and follow the lowest value of the Hydrogen Council for the low, the IEA’s normal cost assumption for the high-cost sensitivity.

In CG H$_{2}$ storage systems, Sens et al. report costs of \$286-\$780/kg H$_{2}$ in 2030 \cite{sens_conditioned_2022}. Their mean value serves as our base case for CG, the lower and upper boundaries as the sensitivity costs.
In stationary PEMFC costs, the European Union’s Clean Hydrogen Partnership (2019) announced target cost declines from $\$_{2019}$1,900/kW$_{\mathrm{e}}$ in 2020 to $\$_{2019}$900/kW$_{\mathrm{e}}$ in 2030 \cite{clean_hydrogen_partnership_hydrogen_2019}. Cigolotti et al. (2021) expect cost drops from $\$_{2021}$2,000-3,500/kW$_{\mathrm{e}}$ to $\$_{2021}$1,200-1,750/kW$_{\mathrm{e}}$ by 2030 due to economies of scale \cite{cigolotti_comprehensive_2021}. We set their mean assumption as our base case. In the sensitivity analysis, the Clean Hydrogen Partnership’s target serves as the low, with Cigolotti et al.’s upper value as the high case.

Uncertainties in the energy storage system comprise whether a big battery park is functional at the altitudes discussed and whether a green H$_{2}$ economy 
will develop until 2030. To find meaningful power system designs for AtLAST, we compare systems with hybrid energy storage against designs without a battery park or without H$_{2}$.

\subsubsection{Line and fuel costs}
\label{sec2-2-3}
For Valley Site power systems, transmission lines need to be built to reach the ATLAST Site. We estimate 43km of newly built 24kV subterranean lines between the generation site and the telescope. Overhead lines are not applicable due to the sulfur-rich environment and harsh weather conditions. We apply costs from a Chilean contractor for subterranean lines plus a 30\% surcharge for the remote and harsh conditions \cite{cype_ingenieros_sa_precio_nodate}. 

For the option of using off-site produced H$_{2}$, we follow the Chilean government’s (2021) assumption of $\$_{2021}$1.05 per kg of green H$_{2}$ produced in 2030, based on analytics by McKinsey \& Company \cite{ministerio_de_energia_gobierno_de_chile_national_2021}. IRENA (2022) expects costs of green H$_{2}$ in Chile of \$1.4/kg H$_{2}$ in 2030, which shall serve as our high sensitivity \cite{international_renewable_energy_agency_global_2022}. Following the work of Tashie-Lewis and Nnabuife (2021), we consider $\$_{2021}$0.50/kg/100km for transporting H$_{2}$ in trucks \cite{tashie-lewis_hydrogen_2021}. We assume to import H$_{2}$ from the Pauna Solar park in María Elena weekly 
\cite{statkraft_statkraft_nodate, cabello_aprobacion_2022}. The trucking distance to the ATLAST Site is 250km, to the Valley Site 200km. All fuel costs are listed in Table \ref{t.fuelcosts}.

With diesel, we are subject to volatile fossil fuel prices. 2022’s peak Chilean diesel price was at \$1220.52/m$^3$, nearly double the previous 4-year average of $\$$566.28/m$^3$ \cite{empresa_nacional_del_petroleo_inversionistas_nodate}. In private communications with the authors, Rystad Energy forecasted the Brent Crude Oil price to decline by 103\% from 2022’s high of \$122.71/b to \$60.40/b in 2030 \cite{empresa_nacional_del_petroleo_inversionistas_nodate}. Applying this ratio to the Chilean diesel price, we result in \$601/m$^3$ in 2030 in our base case. We include 2022’s peak as the high sensitivity case. Shipping of diesel is included in the diesel generator OPEX, see Table \ref{ta.capexopex}. 

\section{Results}
\label{sec3}

The model highRES-AtLAST optimizes total system costs (annualized investment and dispatch costs). The results include the built capacity, dispatch, direct CO$_{2}$e emissions from operation and the total electricity dispatch costs, amongst others. To calculate LCOEs of the system, the latter is divided by the annual demand of 7.7GWh:

\[ \text{LCOE} = \frac{\text{Annualized total electricity system cost}}{\text{Annual demand}}. \]

In the following section, we discuss the base case scenario optimizations. We look at changing results when one energy storage option is excluded and further dive into costs from sensitivity analyses.

\subsection{Base case optimized power systems}
\label{sec3-1}
\begin{figure}
\centering
\includegraphics[width=10.5cm]{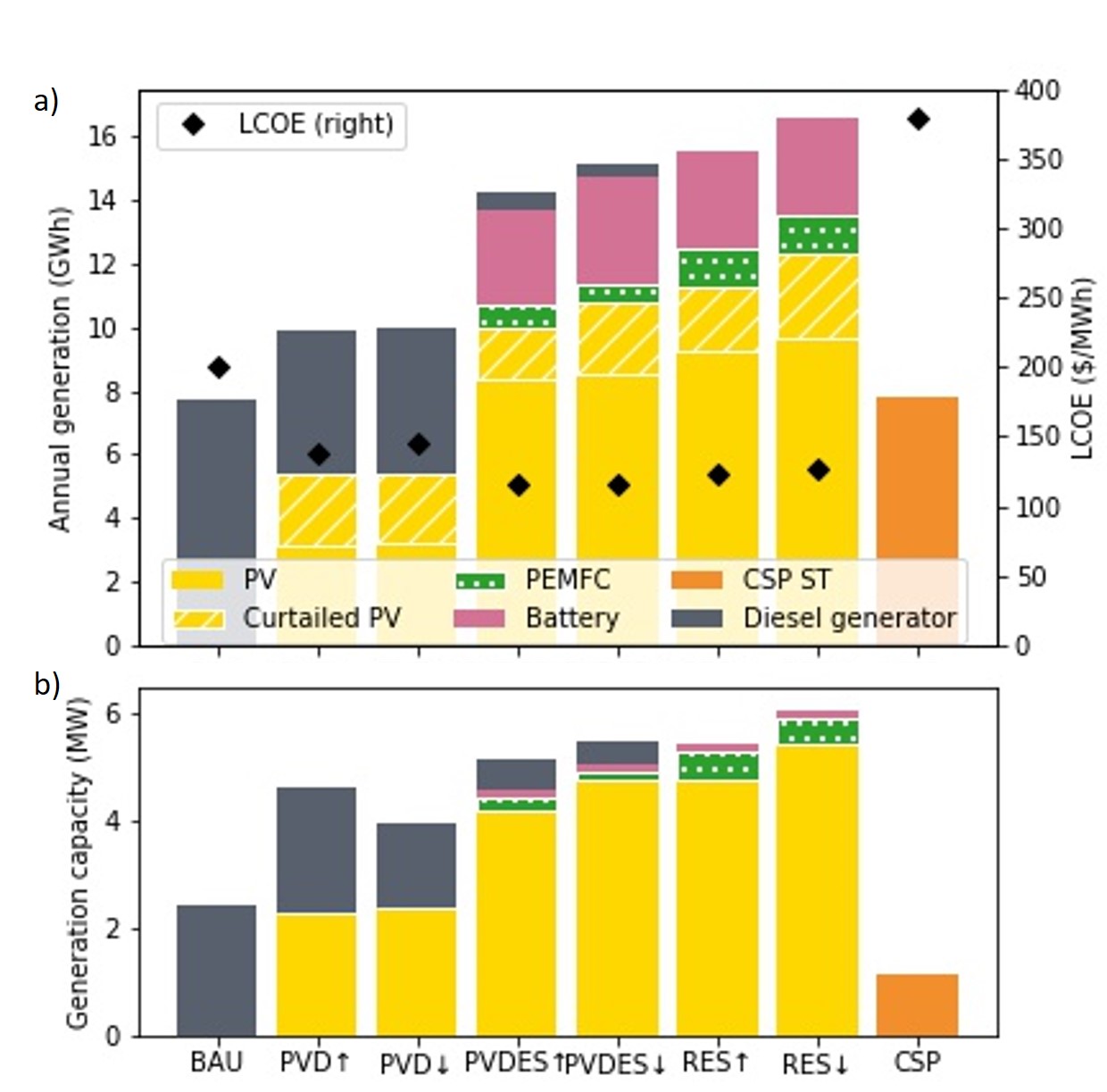}
\caption{Results base case: (a) Annual generation in GWh and LCOEs in \$/MWh, (b) built capacity in MW.\label{fig.resnorm}}
\end{figure}

\begin{figure}
\centering
\includegraphics[width=16.5 cm]{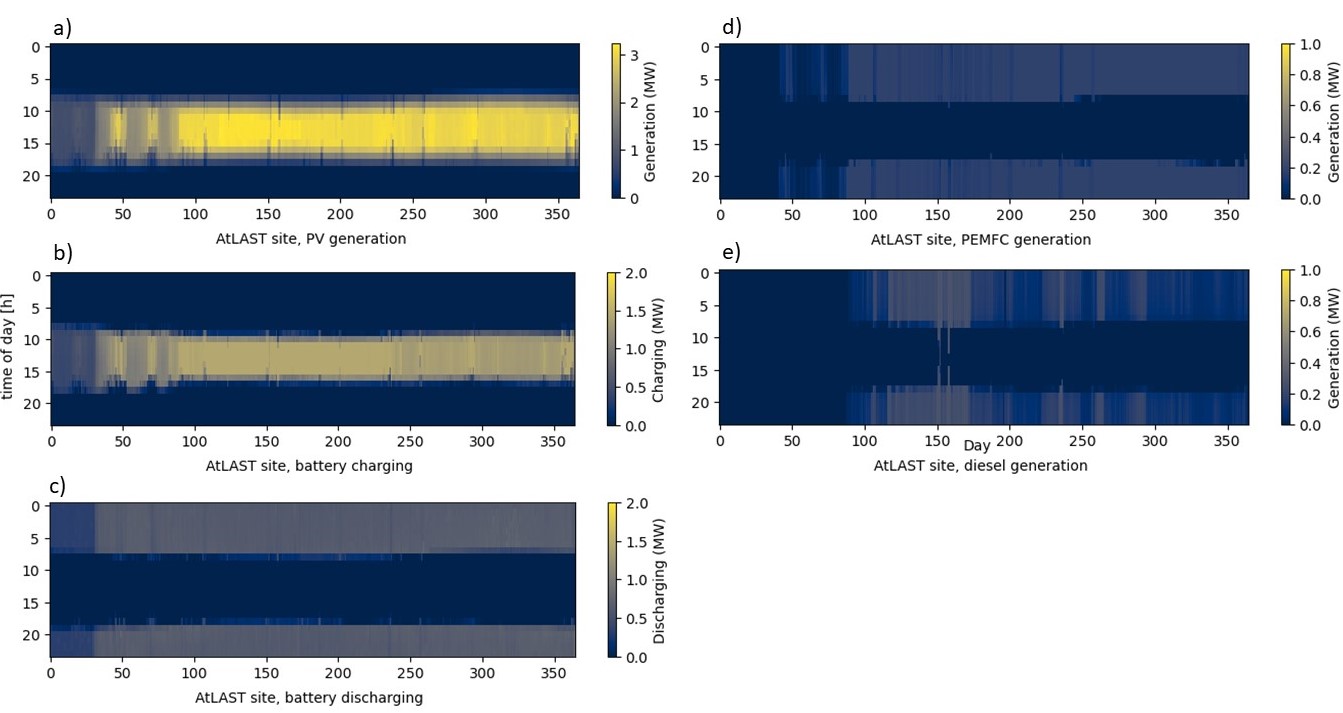}
\caption{Power generated and battery charging/discharging over optimized year, scenario PVDES↑: (a) PV generation, (b) battery charging, (c) battery discharging, (d) PEMFC generation, (e) diesel generation.\label{fig.pvdes}}
\end{figure}

Figure \ref{fig.resnorm} shows the LCOEs, annual generation and built capacity of power generating units of the optimized scenarios, assuming costs as in our base case. See Table \ref{ta.results} in Appendix for an extensive overview of the results.
Scenarios with PV generation achieve LCOEs between \$116.0 and \$144.6/MWh. Employing PV together with energy storage solutions (PVDES and RES) results in lower generation costs than supplying the demand not covered by PV mostly with diesel generators (PVD). The lowest costs are achieved in PVDES, where the power system consists of PV, batteries, H$_{2}$ and diesel for backup. The PVDES and RES systems produce 51-56\% of their H$_{2}$ supply on-site, the rest is imported. When diving into the hourly generation data, we see that batteries and PEMFC generation are used throughout the nights of the simulated year, see Figure \ref{fig.pvdes}. Power generation from diesel is mostly called during periods of less sunny days. The Chilean summer months January to March have enough solar generation and storage capacity to meet the seasonally lower telescope demand, while the autumn and winter months use generation from diesel for cloudy periods. When cutting out diesel generators in the optimization, LCOE values rise by 6-9 percentage points (pp.), including more generation from green H$_{2}$. The CSP scenario results in LCOEs of \$379.5/MWh.

\unskip

To evaluate the direct carbon footprint associated with the energy system in each scenario, we considered that only the diesel generator emits GHG during its operation. Assuming an emission factor of 0.86t CO$_{2}$e/MWh \cite{gaete-morales_assessing_2018}, the direct GHG emissions, measured in CO$_{2}$e, are highest in BAU, see Table \ref{t.emissions}. Pairing diesel with PV reduces the annual emissions by about 40\%, while PVDES scenarios cut 93-95\% of the diesel-only emissions. Scenarios without diesel generators forego any direct emissions.

\begin{table}
\caption{Direct CO$_{2}$e emissions of the base case optimizations.\label{t.emissions}}
\begin{tabular}{lrrrrrrrr}
\toprule
Scenario	& BAU&PVD↑&PVD↓&PVDES↑&PVDES↓&RES↑&RES↓&CSP\\
\midrule
Annual emissions\\
(t CO$_{2}$e/year)&6,624	&3,892&	3,954&465	&323&0&0&0	\\
CO$_{2}$e reduction\\
compared to BAU (\%) &--&41.20\%&
	40.30\%&93.00\%&	95.10\%&	100\%&	100\% &	100\%\\
\bottomrule
\end{tabular}
\end{table}

\subsection{Systems without battery park or without green hydrogen}
\label{sec3-2}
\begin{figure}
\centering
\includegraphics[width=16cm]{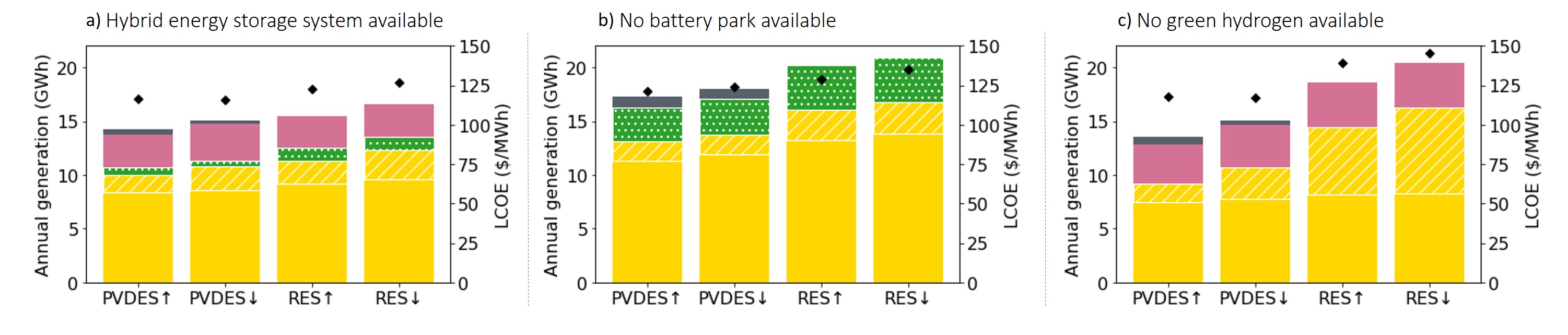}
\caption{Annual generation in GWh and LCOEs in \$/MWh: (a) Base case with hybrid energy storage, (b) base case without battery park, (c) base case without green H$_{2}$.\label{fig.nobatt}}
\end{figure}

Figure \ref{fig.nobatt} compares results from the base case PVDES and RES scenarios without a battery park and without H$_{2}$. In cases without a battery park, resulting power systems mainly consist of PV and green H$_{2}$. LCOEs range \$121-124/MWh with diesel as backup, and \$129-135/MWh without. A small capacity of batteries is built to balance frequency and voltage, assumed at 0.01MWh. The resulting systems in median produce 62\% of the needed H$_{2}$ on-site. In the H$_{2}$-based RES scenarios, an LCOE increase of 6pp. must be taken to reduce the final 13\% of CO$_{2}$e emissions from PVDES. The size of PV parks increases by 27-42\% compared to the base case with batteries, as more PV generation fuels electrolyzers. When cutting out H$_{2}$ as an option, results from the PVDES and RES scenarios change as shown in Figure \ref{fig.nobatt}c. We see setups based on larger PV and battery capacities. The PV capacity in the RES scenarios increases by 28-32\%; with 44-49\% of its generation curtailed. LCOEs lie \textasciitilde\$117/MWh when diesel serves as backup, and \$139-145/MWh without diesel in the system. In the RES scenarios without H$_{2}$, LCOEs increase by 10pp. compared to their PVDES counterpart. This saves the last \textasciitilde8.6\% of CO$_{2}$e emissions which PVDES without H$_{2}$ emits.

\subsection{Sensitivity analyses}
\label{sec3-4}
\begin{figure}
	\centering
	\includegraphics[width=13.5cm]{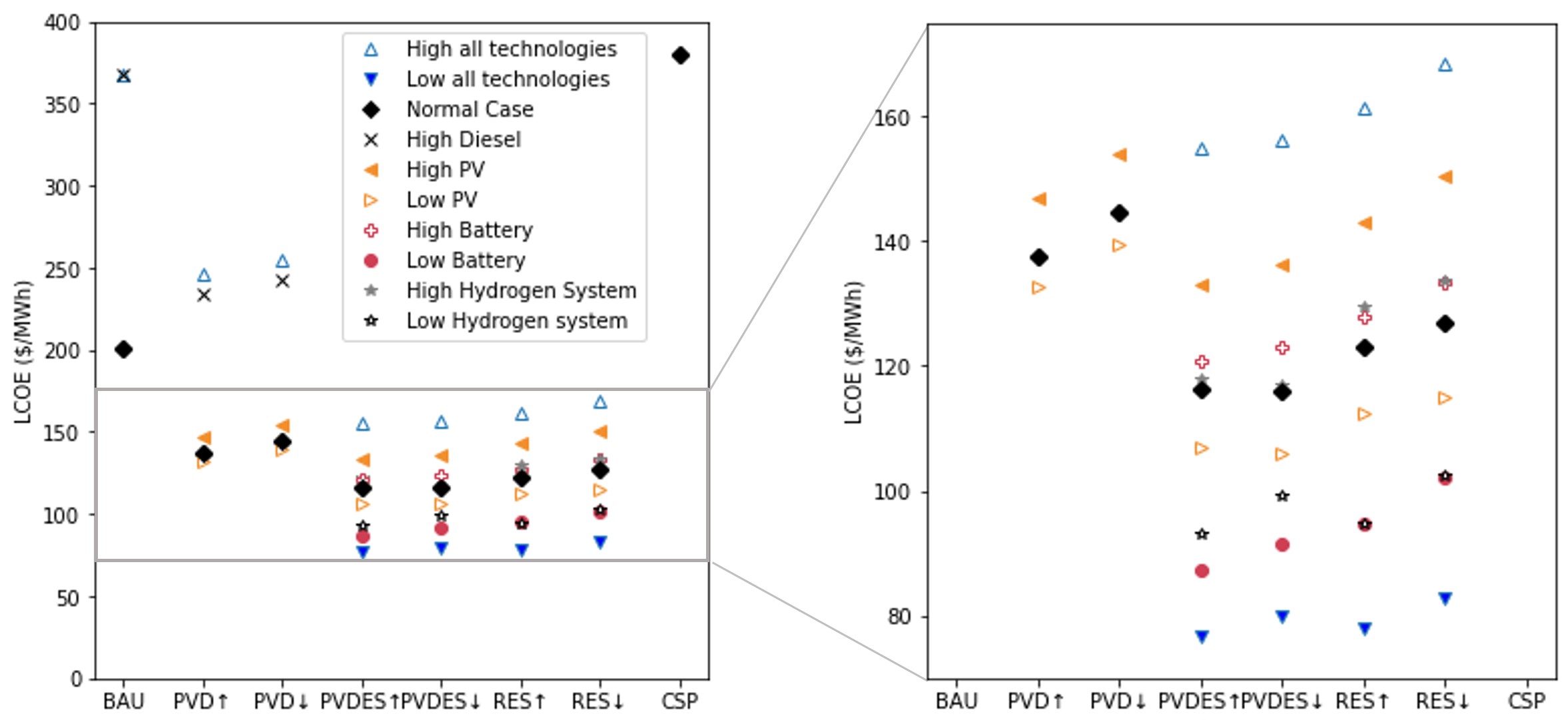}
	\caption{LCOE comparison, base case costs (black diamonds) and sensitivity cost assumptions.\label{fig.lcoe}}
\end{figure}

Optimizing the power system with the sensitivity cost assumptions as in Section \ref{sec2-2} results in varying system setups and costs. The resulting LCOEs are shown in Figure \ref{fig.lcoe}, with the values resulting from the base case costs as black diamonds. In the H$_{2}$ systems high sensitivity, high costs of electrolyzers, fuel cells and imported green H$_{2}$ are applied coincidently. In the low case, low costs of electrolyers and PEMFC are used with base case green H$_{2}$ costs.

Higher fossil fuel costs lead to hefty LCOE jumps of 83\% in the BAU scenario. Similar jumps of 70\% are present in the PVD scenarios with high diesel costs, while PVDES only sees a cost increase of 3\%.
The scenarios with energy storage options, PVDES and RES, respond most to higher PV costs on the upper end. On the lower end, lower battery and lower H$_{2}$ system costs make the biggest difference on them.

\section{Discussion}
\label{sec4}

All scenarios calculated result in systems that can cover the forecasted annual demand of the telescope AtLAST, as all optimizations converge to an optimal solution. Obtained LCOEs in the base case lie \$116-201/MWh, that is below up to a similar range to those of stationary systems for residential demanders discussed in Section \ref{sec1}. Lower generation costs in the power system setups presented are plausible given the cost decline assumptions applied, and pointing to the global solar maximum in the Atacama. Regarding our research questions, we find:
\begin{enumerate}
\item The base case cost assumptions lead to low LCOEs in PVDES and RES scenarios, with most of the generation carried by PV and batteries, next to some H$_{2}$-driven fuel cells. Lowest costs with \$116/MWh are achieved in PVDES, with generators running on diesel used for cloudy days.
\item An all renewable power system as in RES↑ has LCOEs increase by 6\% compared to PVDES↑. This small spike avoids the final 7\% of the BAU scenario’s emissions, equaling 465t of CO$_{2}$e, or 183 European researchers flying roundtrip to AtLAST (see CO$_{2}$e emissions calculation with the Environmental Footprint 3.0 Method in Supplementary Material).
\item	In the sensitivity analyses with upper and lower boundaries of expected costs in 2030, see Section \ref{sec2-2}, a jump above the base case cost value window of \$116-201/MWh presents itself in the diesel-only scenario. When applying the peak Chilean diesel price, system costs increase by 100\% compared to the base case. It is deduced that a power system solely dependent on a fossil source of power generation leads to fickle power generation costs. We propose to employ a system with PV and applicable energy storage to power AtLAST sustainably, both in terms of GHG emissions, but also more robust costs.
\end{enumerate}

GHG emissions related to the operation of the power system in scenarios swerving from the BAU approach reduce CO$_{2}$e-emissions by 40 to 100\%. PEM electrolyzers are preferred over alkaline ones, CSP ST is preferred over CSP PT. Systems that employ H$_{2}$ both use imported and on-site produced H$_{2}$. The import is often used to cover H$_{2}$ demand peaks when high power demand meets cloudy weather.

Both systems without H$_{2}$ or without batteries were simulated to consider cases where these technologies cannot be considered for the telescope. This could e.g. occur in case the electrolyte of batteries does not operate well at the altitude, or in case green H$_{2}$ is not widely available as a fuel in 2030. Both instances result in more usage of diesel generation in PVDES, as seen in Figure \ref{fig.nobatt}b and c. Not having batteries is slightly more expensive than not having H$_{2}$. In RES, more PV capacity is needed to either supply a bigger battery park or electrolyzers. Decarbonifying the battery-rich systems without green H$_{2}$ results in high PV curtailments of up to 40\%. Scenarios relying on PV and storage solutions generally resulted in lower costs than diesel-relying scenarios BAU and PVD, regardless whether batteries, green H$_{2}$ or a combination of the two are used. Low-CO$_{2}$e emitting systems in PVDES need to install less PV and storage capacity than 100\% renewable scenarios, resulting in lower costs and less curtailment.

Comparing PV and CSP systems, CSP presented costs of \$380/MWh, vastly higher than any PV scenario. With CSP costs forecasted as in this paper, this technology is not an economically viable option for AtLAST. To study the feasibility of this solar technology in larger systems, one could include additional demands, like neighboring telescopes or a nearby city, in future case studies.

\subsection{Weather data, climate change reliability}
\label{sec4-2}
The solar capacity factors in this work were based on solely two ERA5 squares of 30km$^2$. With altitude changes from 2,500m to 5,000m, considerable variations in the weather conditions are expected. These cannot be represented within a single square grid, as the irradiation within is considered uniform. To better estimate the solar generation at the AtLAST and Valley Site, future work should consider the use of a higher resolution database.

For solar generation, we assumed the weather conditions for 2030 as similar to 2020. While estimations of solar variation expect the Chajnator area to undergo no high changes in irradiation until 2065 \cite{schoniger_making_2021}, these small changes can be propagated during the system’s lifetime and affect its reliability. To design a reliable power system and reduce the uncertainty associated to weather data, a future climate scenario database could be considered in further investigations. 

\subsection{High altitude considerations}
\label{sec4-3}
This work considered both power systems next to the new telescope at an altiutude of \textasciitilde5,000m and systems down in the valley at \textasciitilde2,500m, connected to the telescope via medium-voltage lines. The altitude made little difference to the resulting system costs, so that for example PVDES↑ and PVDES↓ both had LCOEs of \$116/MWh. While power systems at the Valley Site consider the costs of the line going up to the Chajnator plateau, the derating factors and higher costs in imported H$_{2}$ increased costs at the high site. While cost-wise, building on either side is reasonable, we cannot be certain that the components considered can operate at the high altitude, as research in this area is lacking. Testing on-site or in laboratory recreating the local conditions will help to answer this.

Further, snowfall needs to be considered at the AtLAST Site, as snow layers can reduce PV performance \cite{santhakumari_review_2019}. To account for this, higher OPEX due to snow handling could be introduced. Moreover, one could find best locations for PV arrays, that is sites with orography leading to less snow accumulation on the panels.

\subsection{Expanding the optimization: Life cycle analysis and energy communities}
\label{sec4-4}
GHG emissions as presented in this paper include direct emissions only. When looking at the carbon footprint over the lifecycle of system components, indirect emissions from raw material sourcing, production, transport and end-of-life treatments should be considered as well, as RES incur indirect up- and downstream GHG emissions. This analysis is the topic of an upcoming paper in preparation by the authors. In this next work, we will include a life cycle assessment within highRES-AtLAST. Using a multi-objective optimization on both costs and lifecycle GHG emissions can generate a more profound answer to which systems induces lowest CO$_{2}$e-emissions.

The systems presented are built to serve a single telescope with an annual demand of 7.7GWh. In proximity to AtLAST on the Chajnator plateau, eleven other telescopes are operating and could be integrated into a joint power system. This would not only reduce costs due to bigger scale, but also require less area and materials per generated Wh. A Valley Site power system, powering multiple telescopes on Chajnator, would only require one power line to the plateau. One could moreover create an energy community with nearby localities, supplying excess power to locals.

\section{Conclusions}
\label{sec5}

The authors found lowest LCOEs of \$116/MWh for the telescope AtLAST with a power system employing a combination of PV, batteries, imported and on-site produced green H$_{2}$ with CG storage and PEMFC, and backup diesel. Compared to a business-as-usual setup with diesel generators only, systems based on PV present 31-42\% lower costs of power generation and more robustness regarding cost sensitivities. They save 40-100\% in direct GHG emissions compared to the BAU scenario. Systems running on 100\% RES with hybrid energy storage have 6-9\% higher costs than those with additional diesel as backup. Setups where PV was used with only one main energy storage option, either batteries or H$_{2}$, result in LCOE jumps of 1-12\%.

Synergies by building a centralized power system for the multitude of telescopes on the Chajnantor plateau could lead to lower costs, e.g. in maintenance and cabling. Further research is needed to conclude whether the systems presented here can be employed at the altitudes discussed and what indirect emissions they entail. The open-source system design optimization of this work can also be applied for other (remote) facilities and thus serve as a lighthouse to astronomical observatories and off-grid stationary applications that would like to shift towards more sustainable power supply.

\section*{Credits}
Conceptualization, I.V., G.V.V., M.Z., and S.S.; methodology, I.V. and G.V.V.; software, G.V.V.; investigation, I.V. and G.V.V.; data curation, I.V. and G.V.V.; writing—original draft preparation, I.V.; writing—review and editing, G.V.V., M.Z. and S.S.; visualization, I.V.; supervision, M.Z. and S.S.; project administration, S.S.; funding acquisition, S.S. All authors have read and agreed to the published version of the manuscript.

\section*{Funding}
This project has received funding from the European Union’s Horizon 2020 research and innovation programme under grant agreement No. 951815.

\section*{Data availability}
The highRES-AtLAST model is published here: https://github.com/highRES-model/highRES-AtLAST. Publicly available hourly solar irradiation time series from ERA5 used in this study can be found here: https://www.ecmwf.int/en/ forecasts/datasets/reanalysis-datasets/era5. Restrictions apply to the availability of the demand dataset, which was upscaled from a dataset provided by APEX, not publicly available. A dummy demand dataset to obtain similar results to the ones presented in this work is openly available within the highRES-AtLAST model.
\section*{Acknowledgements}
We want to thank APEX for their kind collaboration. Special thanks goes to our astronomy colleagues Claudia Cicone, Carlos De Breuck and Tony Mroczkowski for the cooperation in this interdisciplinary project.


\appendix
\section*{Appendix}\label{seca}
\setcounter{table}{0}
\renewcommand{\thetable}{A.\arabic{table}}

\begin{table}[ht!]
\caption{CAPEX and OPEX estimations in 2030, real2022-US\$ values.\label{ta.capexopex}}
	\begin{tabular}{lrrrrrr}
	\toprule
			Cost component	& CAPEX	& Unit & Fixed OPEX & Unit/year	& Variable OPEX & Unit\\
			\midrule
\makecell[l]{Monofacial monocrys-\\talline PV} & 523 \cite{nama_facility_indice_2020, vartiainen_impact_2020} & \$/kW$_{\mathrm{p}}$ & 1\% \cite{vartiainen_impact_2020} & \makecell{\% of CAPEX} &--&--\\
CSP Solar Field – PT / ST & 209/148\cite{turchi_csp_2019} & \$/m$^{2}$ & 57.35 \cite{turchi_csp_2019} & \$/kWh & 3.48/3.04 \cite{turchi_csp_2019} & \multirow{4}{*}{\$/MWh}\\
Diesel generator & 495 \cite{comision_nacional_de_energia_informe_2020} &\$/kW &--&--& \makecell[r]{23.5 (10 \cite{comision_nacional_de_energia_informe_2020} \\ +0.5\$/l for ship-\\ping (assumed))} & \\
Li-Ion batteries & 262 \cite{cole_cost_2021, mongird_2020_2020} & \$/kWh & 0.43\% \cite{cole_cost_2021} & \makecell{\% of CAPEX} &0.5664 \cite{cole_cost_2021} & \\
Electrolyzer - Alkaline / PEM&434/483 \cite{cole_cost_2021}& \$/kW$_{\mathrm{e}}$&14.1\cite{cole_cost_2021} & \$/kW$_{\mathrm{e}}$ & 0.9394 \cite{cole_cost_2021} & \$/MWh$_{\mathrm{e}}$\\
CG H$_{2}$ storage & 598 \cite{sens_conditioned_2022} & \makecell{\$/kg H$_{2}$\\stored} & 2.1\% \cite{cole_cost_2021} & \makecell{\% of CAPEX} &--&--\\
PEMFC & 1,581 \cite{cigolotti_comprehensive_2021} & \$/kW$_{\mathrm{e}}$ & 14.1 \cite{cole_cost_2021} &\$/kW$_{\mathrm{e}}$&0.9394 \cite{mongird_2020_2020} & \$/MWh$_{\mathrm{e}}$\\
\makecell[l]{Subterranean power\\lines, 24kV} & 23,599 \cite{cype_ingenieros_sa_precio_nodate}&\$/km & \makecell[r]{0.5\% \\(assumed)} & \makecell{\% of CAPEX}&--&--\\
			\bottomrule
	\end{tabular}
\end{table}

\begin{table}[ht!]
\caption{Results of base case cost optimization.\label{ta.results}}
		\begin{tabular}{lrrrrrrrrr}
			\toprule
			Scenario&Unit	& BAU	& PVD↑	& PVD↓ & PVDES↑& PVDES↓	& RES↑ & RES↓& CSP\\
			\midrule
			Annualized total \\electricity system cost& k\$&1,544.7&1,059.9&1,112.8&894.9&893.3&946.1&978.0&2,921.6\\
\textbf{Capacity}\\
Diesel&\multirow{2}{*}{MW}&2.40&2.34&1.58&0.58&0.43&--&--&\multirow{7}{*}{--}\\
PV&&\multirow{8}{*}{--}&2.27&2.38&4.20&4.76&4.75&5.44&\\
Alkaline electrolyzer& \multirow{2}{*}{MW$_{{\mathrm{e}}}$}&&\multirow{7}{*}{--}&\multirow{7}{*}{--}&0.18&0.34&0.22&0.59&\\
PEM electrolyzer& &&&&0.68&0.39&1.01&0.91&\\
CG storage& kg H$_{2}$&&&&263.1&159.8&1,048.7&1,399.3&\\
PEMFC& MW$_{{\mathrm{e}}}$&&&&0.22&0.16&0.54&0.49&\\
Li-ion battery& MWh&&&&15.1&14.2&15.2&12.4&\\
CSP solar field& \multirow{2}{*}{MW}&&&&\multirow{2}{*}{--}&\multirow{2}{*}{--}&\multirow{2}{*}{--}&\multirow{2}{*}{--}&12.2\\
CSP ST&&&&&&&&&1.14\\
\textbf{Generation}&annual&&&&&&\\
Diesel& \multirow{5}{*}{MWh}&7,697.6&4,523.5&4,594.9&540.8&375.2&--&--&\multirow{5}{*}{--}\\
PV& &\multirow{7}{*}{--}&3,174.2&3,203.4&8,404.0&8,544.4&9,217.7&9,627.1&\\
Thereof curtailed& &&2,223.9&2,183.5&1,570.7&2,254.4&2,065.7&2,700.4&\\
PEMFC& &&\multirow{5}{*}{--}&\multirow{5}{*}{--}&735.2&534.9&1,204.7&1,201.9&\\
Li-ion battery& &&&&2,993.7&3,396.4&3,015.9&3,052.4&\\
CSP solar field& \multirow{2}{*}{MWh$_{{\mathrm{th}}}$}&&&&\multirow{3}{*}{--}&\multirow{3}{*}{--}&\multirow{3}{*}{--}&\multirow{3}{*}{--}&20,668.6\\
Thereof curtailed& &&&&&&&&2,815.0\\
CSP ST& MWh&&&&&&&&7,798.1\\
			\bottomrule
		\end{tabular}
\end{table}


\end{document}